\pdfoutput=1
\PassOptionsToPackage{table,xcdraw}{xcolor}
\documentclass[11pt]{article}

\usepackage{acl}

\usepackage{times}
\usepackage{latexsym}
\usepackage{amsmath}
\usepackage{graphicx}
\usepackage{multirow}
\usepackage{amssymb}
\usepackage{enumerate}
\usepackage{tcolorbox}

\usepackage[T1]{fontenc}

\usepackage[utf8]{inputenc}

\usepackage{microtype}

\usepackage{inconsolata}

\usepackage{graphicx}

\title{Herd Behavior: Investigating Peer Influence in LLM-based Multi-Agent Systems}

\author{Young-Min Cho, Sharath Chandra Guntuku, Lyle Ungar \\
  University of Pennsylvania \\
  \texttt{jch0@seas.upenn.edu}}

\begin{document}
\maketitle
\begin{abstract}
Recent advancements in Large Language Models (LLMs) have enabled the emergence of multi-agent systems where LLMs interact, collaborate, and make decisions in shared environments. While individual model behavior has been extensively studied, the dynamics of peer influence in such systems remain underexplored. In this paper, we investigate \textbf{herd behavior}, the tendency of agents to align their outputs with those of their peers, within LLM-based multi-agent interactions. We present a series of controlled experiments that reveal how herd behaviors are shaped by multiple factors. First, we show that the gap between self-confidence and perceived confidence in peers significantly impacts an agent’s likelihood to conform. Second, we find that the format in which peer information is presented plays a critical role in modulating the strength of herd behavior. Finally, we demonstrate that the degree of herd behavior can be systematically controlled, and that appropriately calibrated herd tendencies can enhance collaborative outcomes. These findings offer new insights into the social dynamics of LLM-based systems and open pathways for designing more effective and adaptive multi-agent collaboration frameworks\footnote{Code and data will be released in the camera-ready.}.
\end{abstract}

\section{Introduction}
Herd behavior refers to the phenomenon of individuals in a group to mimic the actions, decisions, or behaviors of a larger group, often disregarding their own analysis or instincts \cite{banerjee1992simple, bikhchandani1992theory}. Humans often adjust their behavior in response to observing peers, aligning their decisions towards perceived group consensus \cite{raafat2009herding, muchnik2013social}. This human tendency raises questions about whether similar dynamics emerge in artificial intelligence. In Large Language Model (LLM)-based multi-agent systems (MAS), multiple autonomous agents powered by LLMs interact and reason collectively, creating fertile ground for social behaviors such as conformity to emerge \cite{guo2024large, park2023generative}. Understanding whether and how these agents exhibit herd behavior is crucial for evaluating the robustness, diversity, and effectiveness of collective decision-making.

\begin{figure}[t]
    \centering
    \includegraphics[width=\linewidth]{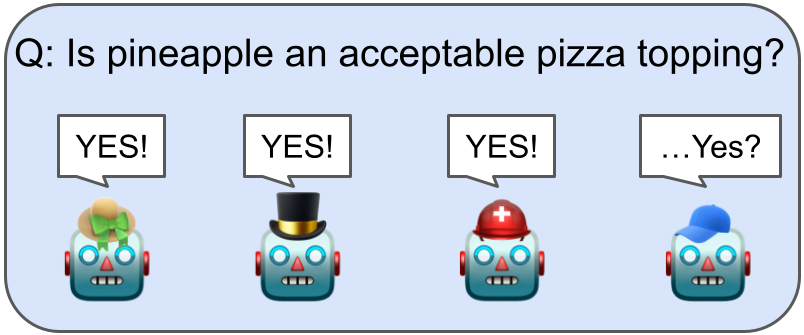}
    \caption{An example of herd behavior: Even when uncertain, individuals tend to follow the crowd, sometimes against their own judgment.}
    \label{fig:spirit}
\end{figure}

Herd behavior in LLM-based MAS can be a double-edged sword. On one hand, convergence towards a group consensus can streamline decision-making, reduce conflict, and enhance coordination, particularly in scenarios where agreement or collective confidence is desirable \cite{guo2024large}. It can also serve as a mechanism for amplifying strong signals or leveraging collective intelligence, allowing agents to compensate for individual uncertainty by incorporating peer input \cite{liu2024groupdebate, du2023improving}. On the other hand, excessive conformity can suppress diversity of thought, lead to premature consensus, and propagate errors if initial signals are flawed \cite{cho2024roundtable, weng2025we}. Such blind alignment may reduce the system’s robustness, hinder exploration of alternative solutions, and make the collective more susceptible to cascading failures \cite{wu2025hidden, zhu2024conformity}. Understanding when herd behavior is beneficial and when it is detrimental is essential for building trustworthy, adaptive, and resilient multi-agent LLM systems.

However, the mechanisms underlying the emergence of herd behavior, as well as the factors that modulate its intensity, remain understudied in the context of LLM-driven multi-agent collaboration. Understanding and intentionally managing herd behaviors within multi-agent collaborations is crucial. 

In this study, we design a set of controlled experiments using LLM-based agents to investigate herd behaviors in MAS. We manipulate key variables such as agents’ self-confidence, perceived peer confidence, and the format of peer information presentation to systematically observe their influence on conformity behavior. By quantifying alignment patterns and measuring task outcomes under different conditions, we uncover the mechanisms behind herd tendencies and explore how they can be tuned to optimize collaboration quality. We find that flip rates peak when agents have very low self-confidence and perceive peers as confident (Figure \ref{fig:confidence}), with the most persuading peer answer driving the strongest herding (0.48 avg. flip rate) but also reducing accuracy on factual tasks (Table \ref{tbl:accuracy}). Format of peer information also significantly impact the herd behavior, where using the combination of factors that amplify herding yields the highest flip rate (0.63) and group accuracy (0.29). In contrast, prompt-based controls have minimal effect (Table \ref{tbl:control}).

Our experiments provide the following contributions:
\begin{enumerate}
    \item We find that herd behavior in LLM-based agents is primarily driven by the relationship between an agent’s self-confidence and its perceived confidence in peers. In particular, larger gaps between these two measures significantly increase the likelihood of conformity.

    \item We show that the presentation format of peer responses critically affects the degree of herd behavior. Notably, placing disagreeing opinions before agreeing ones amplifies conformity, suggesting that ordering and framing effects shape social influence among agents.

    \item We demonstrate that herd behavior can be systematically tuned, and that appropriate calibration of conformity levels can enhance the effectiveness of multi-agent collaboration, offering design principles for future adaptive MAS systems.
\end{enumerate}

\section{Preliminaries}
In this section, we introduce the preliminaries of the experiments.

\paragraph{Problem setting.} In a multi-agent collaboration, each agent $ a_i \in A $ is prompted with a question $ q $ and provides a response $ r_i \in R $, where:

\begin{itemize}
  \item $ A = \{a_1, a_2, \ldots, a_{|A|}\} $ is the set of agents involved in the collaboration,
  \item $ R = \{r_1, r_2, \ldots, r_{|A|}\} $ is the corresponding set of responses, where $ r_i $ is the response from agent $ a_i $.
\end{itemize}

All agents share the same generation distribution $P_\tau(\cdot \mid C)$, which is conditioned on the context $C$ and modulated by temperature $\tau$. The context $C$ for each agent includes the question $q$ and optionally the responses of the other agents, denoted as $R_{-i} = \{ r_j \mid j \neq i, a_j \in A \}$. Each agent selects the response with the highest probability under this distribution. Agents do not have an external memory module.

For simplicity, all questions are multiple choice questions, where $r \in \mathcal{R} = \{ \text{A}, \text{B}, \text{C}, \dots \}$ is one of the discrete candidate responses, and $ \mathcal{R} $ is the set of all candidate responses for question $ q $.


\paragraph{Definition 1: Confidence.} Following the works of \citealt{xiao2019quantifying}, we define an agent's confidence (preference) in its response to a question as the probability assigned by the generation distribution $P(r \mid C)$. Since the responses are fixed categorical choices, we treat each $r$ as a single-token label, and define the confidence as:
\[
P(r \mid C) = \frac{\exp(z_r)}{\sum_{r' \in \mathcal{R}} \exp(z_{r'})},
\]
where $ z_r $ is the unnormalized logit score for choice $ r $. The higher the probability assigned to a response $ r $, the more confident the agent is in its correctness.

\paragraph{Definition 2: Preference Update.}
We define how an agent's response preference changes when peer information is introduced. Given a question $q$:

\begin{itemize}
  \item The \textbf{original response} of the agent, based solely on the question, is defined as:
  \[
  r' = \arg\max_{r \in \mathcal{R}} P(r \mid q)
  \]
  \item The \textbf{revised response}, incorporating peer information, is defined as:
  \[
  r^h = \arg\max_{r \in \mathcal{R}} P(r \mid q, R_{-i})
  \]
\end{itemize}

This formulation captures how the presence of other agents' responses $R_{-i}$ can influence an agent's selected answer.

\paragraph{Definition 3: Herd Behavior.} Following the works of \citealt{laban2023you}, we define herd behavior as the tendency of an agent to change its initial decision after observing or interacting with others. Formally, we define the herd behavior of an agent $a_i$ on question $q_k$ as a binary indicator:
\[
\mathbb{I}_{\text{flip}}(a_i, q_k) = 
\begin{cases}
1, & \text{if } r'_{i,k} \neq r_{i,k}^h \\
0, & \text{otherwise}
\end{cases}
\]

We define the \textbf{flip rate} as the average fraction of agents who changed their answers, aggregated across all questions:
\[
\text{Flip Rate} = \frac{1}{|Q| \cdot |A|} \sum_{q_k \in Q} \sum_{a_i \in A} \mathbb{I}_{\text{flip}}(a_i, q_k)
\]
where $Q$ is the set of questions. A higher flip rate indicates a stronger degree of herd behavior.

\section{Self and Perceived Confidence - Primary Driver of Herd Behavior}

Herd behavior in human society is influenced by multiple factors, with numerous studies indicating that confidence plays a central role in driving this phenomenon. Studies in behavioral economics and psychology have shown that confident individuals can disproportionately influence group decisions, especially when others are uncertain \cite{zarnoth1997social, bang2017confidence, fu2017confidence}. In group settings, individuals often defer to those who express higher certainty, regardless of accuracy \cite{pescetelli2021confidence, moussaid2013social}. 

Inspired by previous studies, we explore how confidence influences agents’ tendency to exhibit herd behavior in a MAS setting. Specifically, we categorize confidence into two levels: \textbf{self-confidence} and \textbf{perceived confidence}. Self-confidence refers to how certain an agent is about its own original response, while perceived confidence refers to how confident the agent perceives its peers to be in their original responses. We hypothesize that lower self-confidence, combined with higher perceived confidence in peers, leads to stronger herd behavior.

\subsection{Experiment Setting}
\label{sec:confidence_setting}
To examine the effects of self-confidence and perceived confidence on herd behavior, we adopt a minimal MAS configuration involving only two agents, $a_i$ and $a_j$. This simplification ensures that each agent interacts with only one peer, allowing for clearer attribution of behavioral changes.

From the agent’s original distribution $P(r \mid q)$ over possible responses to question $q$, we manually select one of four types of responses to serve as the peer’s opinion $r_j$:

\noindent \textbf{• 1st}: The most probable response, which coincides with the agent’s original response $r_i$.

\noindent \textbf{• 2nd}: The second most probable response, chosen to represent a highly persuasive alternative from the agent’s perspective.

\noindent \textbf{• rnd}: A randomly sampled response from the distribution $P(r \mid q)$.

\noindent \textbf{• last}: The least probable response, assumed to be the least persuasive to the agent.

\begin{table}[t]
\resizebox{\linewidth}{!}{%
\begin{tabular}{llll}
\hline
\textbf{Type}                         & \textbf{Benchmark}                                                                                           & \begin{tabular}[c]{@{}l@{}}\textbf{Number of }\\ \textbf{Questions}\end{tabular} & \begin{tabular}[c]{@{}l@{}}\textbf{Avg. Number} \\ \textbf{of Choices}\end{tabular} \\ \hline
\multirow{3}{*}{Factual}     & \begin{tabular}[c]{@{}l@{}}MMLU-Pro\\ \cite{wang2024mmlu}\end{tabular}             & 12,032                                                         & 9.47                                                              \\
                             & \begin{tabular}[c]{@{}l@{}}GPQA-Diamond\\ \cite{rein2024gpqa}\end{tabular}         & 198                                                            & 4.00                                                              \\
                             & \begin{tabular}[c]{@{}l@{}}ARC-Challenge\\ \cite{clark2018think}\end{tabular}      & 1,172                                                          & 4.00                                                              \\ \hline
\multirow{3}{*}{Opinionated} & \begin{tabular}[c]{@{}l@{}}OpinionQA\\ \cite{santurkar2023whose}\end{tabular}      & 1,506                                                          & 3.24                                                              \\
                             & \begin{tabular}[c]{@{}l@{}}GlobalOpinionQA\\ \cite{durmus2023towards}\end{tabular} & 2,555                                                          & 4.09                                                              \\
                             & \begin{tabular}[c]{@{}l@{}}SOCIAL IQA\\ \cite{sap2019socialiqa}\end{tabular}       & 1,954                                                          & 3.00                                                              \\ \hline
\end{tabular}
}
\caption{Basic statistics of the benchmarks used in our experiments.}
\label{tbl:benchmark}
\end{table}

Given a question $q \in Q$ and the selected peer opinion $r_j$, the agent generates a revised response $r^h = \arg\max_{r \in \mathcal{R}} P(r \mid q, r_j)$. We then compute the flip rate across all questions, analyzing how the strength of herd behavior related with the agent’s self-confidence $P(r_i \mid q)$ and the perceived confidence $P(r_j \mid q)$.

Additionally, we examine varying degrees of perceived confidence based on the peer’s persona. By manipulating factors such as education level (\textit{graduate degree, college degree, high school diploma}), social hierarchy (\textit{employer vs. employee}), and domain expertise (\textit{in-domain vs. out-of-domain})\footnote{Only MMLU-Pro and GPQA-Diamond contain domain-specific questions. We label a peer as in-domain if their provided expertise matches the question’s domain.}, we investigate how these factors impact the strength of herd behavior. These experiments are performed with \textit{2nd} response type for strongest signal.

\begin{table*}[t]
\resizebox{\textwidth}{!}{%
\begin{tabular}{llllllll}
\hline
\multirow{2}{*}{\textbf{Peer Condition}} & \multicolumn{3}{c}{\textbf{Factual}}                               & \multicolumn{3}{c}{\textbf{Opinionated}}                            & \multirow{2}{*}{\textbf{Average}} \\
                               & \textbf{MMLU-Pro} & \textbf{GPQA-Diamond} & \textbf{ARC-Challenge} & \textbf{OpinionQA} & \textbf{GlobalOpinionQA} & \textbf{SOCIAL IQA} &                                   \\ \hline
1st                            & 0.03              & 0.05                  & 0.01                   & 0.01               & 0.01                     & 0.02                & 0.03                              \\
2nd                            & \textbf{0.51*}    & \textbf{0.58*}        & \textbf{0.09*}         & \textbf{0.61*}     & \textbf{0.69*}           & \textbf{0.16*}      & \textbf{0.48*}                    \\
rnd                            & 0.31              & 0.40                   & 0.04                   & 0.55               & 0.60                      & 0.09                & 0.33                              \\
last                           & 0.25              & 0.37                  & 0.04                   & 0.52               & 0.56                     & 0.09                & 0.29                              \\ \hline
Graduate Degree                & \textbf{0.50*}    & \textbf{0.56*}        & \textbf{0.08}          & \textbf{0.76*}     & \textbf{0.83*}           & \textbf{0.15*}      & \textbf{0.51*}                    \\
College Degree                 & 0.47              & 0.49                  & 0.08          & 0.74               & 0.83                     & 0.14                & 0.48                              \\
High School Diploma            & 0.44              & 0.48                  & 0.07                   & 0.71               & 0.79                     & 0.14                & 0.46                              \\ \hline
Employer                       & \textbf{0.57*}    & 0.71                  & \textbf{0.10}                    & 0.71               & 0.77                     & \textbf{0.20*}      & \textbf{0.54*}                    \\
Employee                       & 0.53              & 0.71                  & 0.09                   & \textbf{0.74*}     & \textbf{0.79*}           & 0.17                & 0.52                              \\ \hline
In-Domain                      & \textbf{0.55*}    & \textbf{0.72*}        & -                      & -                  & -                        & -                   & \textbf{0.55*}                    \\
Out-Of-Domain                  & 0.48              & 0.46                  & -                      & -                  & -                        & -                   & 0.48                              \\ \hline
\end{tabular}
}
\caption{Flip rates across different peer conditions to evaluate the impact of perceived confidence on herd behavior. Bolded values represent the highest flip rate within each group, indicating the strongest herd influence. Asterisks (*) denote statistical significance (p < 0.05) based on paired t-tests within each group.}
\label{tbl:confidence}
\end{table*}

\begin{figure}[t]
    \centering
    \includegraphics[width=\linewidth]{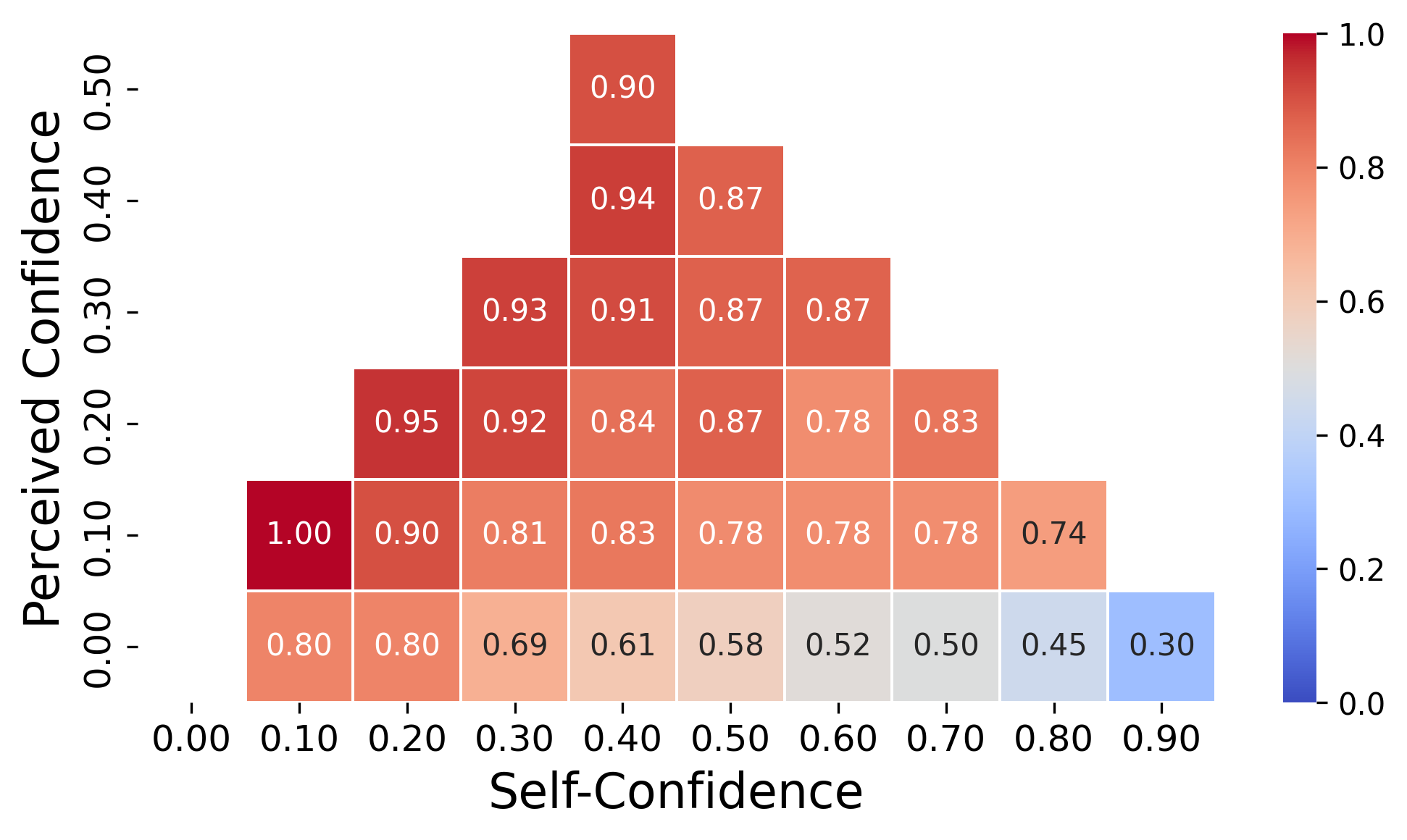}
    \caption{Flip rate across varying levels of self-confidence and perceived confidence. The experiment includes all benchmarks under the \textit{2nd, rnd,} and \textit{last} peer conditions. Lower self-confidence or higher perceived confidence corresponds to stronger herd behavior.}
    \label{fig:confidence}
\end{figure}

\subsection{Dataset}
\label{sec:confidence_dataset}
We select six multiple choice benchmarks to ensure the generalizability of our experiments. They cover both factual and opinionated questions, since real-world decision-making often involves a mix of objective knowledge and subjective judgment. While factual questions have gold answers, opinionated questions do not. The basic statistics of the selected benchmarks are shown in Table \ref{tbl:benchmark}.

\subsection{Results}

\paragraph{Confidence-Driven Herding} Figure \ref{fig:confidence} shows the flip rate under varying levels of self-confidence and perceived confidence, averaged across all benchmarks using \textit{2nd, rnd} and \textit{last} peer conditions. The heatmap reveals a clear pattern: flip rates are highest when self-confidence is low and perceived peer confidence is high, indicating stronger herd behavior in such conditions. As self-confidence increases, individuals become less likely to switch their answers, even when peers appear confident. Conversely, when perceived confidence from peers increases, individuals with low self-confidence are more prone to change their responses. These findings highlight the significant role that both internal certainty and social influence play in shaping decision-making behavior.

\paragraph{Peer Influence Dynamics} Table \ref{tbl:confidence} and Table \ref{tbl:accuracy} examine how perceived confidence from different peer conditions influences the strength of herd behavior. Table \ref{tbl:confidence} shows that \textit{2nd}, the second most probably response consistently results in the highest flip rates across benchmarks, indicating strong tendency to peer influence. Educational background, social hierarchy, and domain relevance also modulate flip rates, where peer's persona with \textit{graduate degree}, \textit{employer} or \textit{in-domain} expertise caused the stronger herd tendencies. Interestingly, \textit{employer} as peer caused weaker herd behavior in opinionated benchmarks, which indicates opposite effect on subjective decisions with social hierarchy.

\paragraph{Herding and Accuracy} Table \ref{tbl:accuracy} evaluates the accuracy of the revised responses after exposure to peer input. Interestingly, the 2nd peer condition, which induces the strongest herd effect, also leads to a statistically significant drop in accuracy compared to the original response, particularly on factual benchmarks like MMLU-Pro and ARC-Challenge. This suggests that following a confident peer does not always yield better outcomes—in some cases, it may degrade accuracy. 

\begin{figure*}[t]
    \centering
    \includegraphics[width=\textwidth]{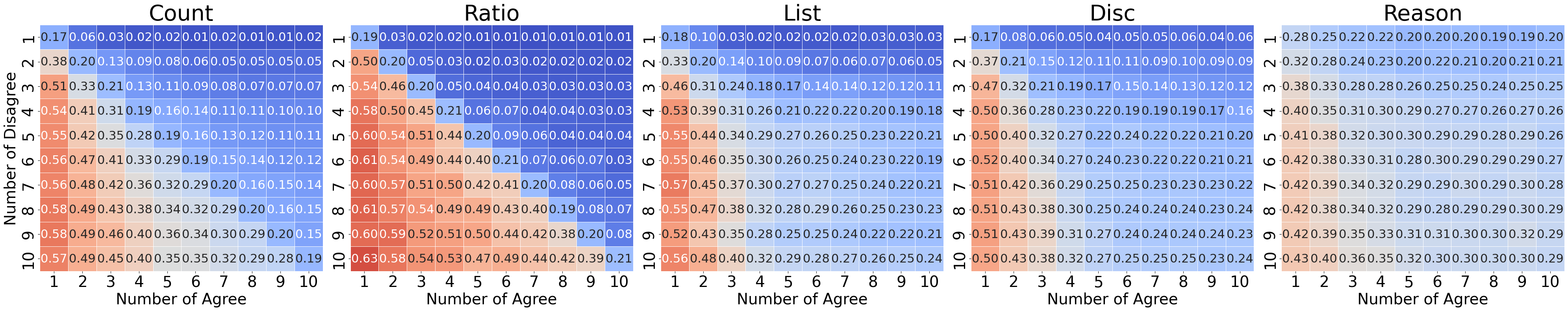}
    \caption{Comparison of average flip rates across five presentation formats. Each heatmap shows the average flip rate based on different combinations of agreeing and disagreeing peers. The x-axis represents the number of agreeing agents, and the y-axis represents the number of disagreeing agents. Higher flip rates are shown in red, while lower rates are shown in blue.}
    \label{fig:format}
\end{figure*}

\begin{figure}[t]
    \centering
    \includegraphics[width=\linewidth]{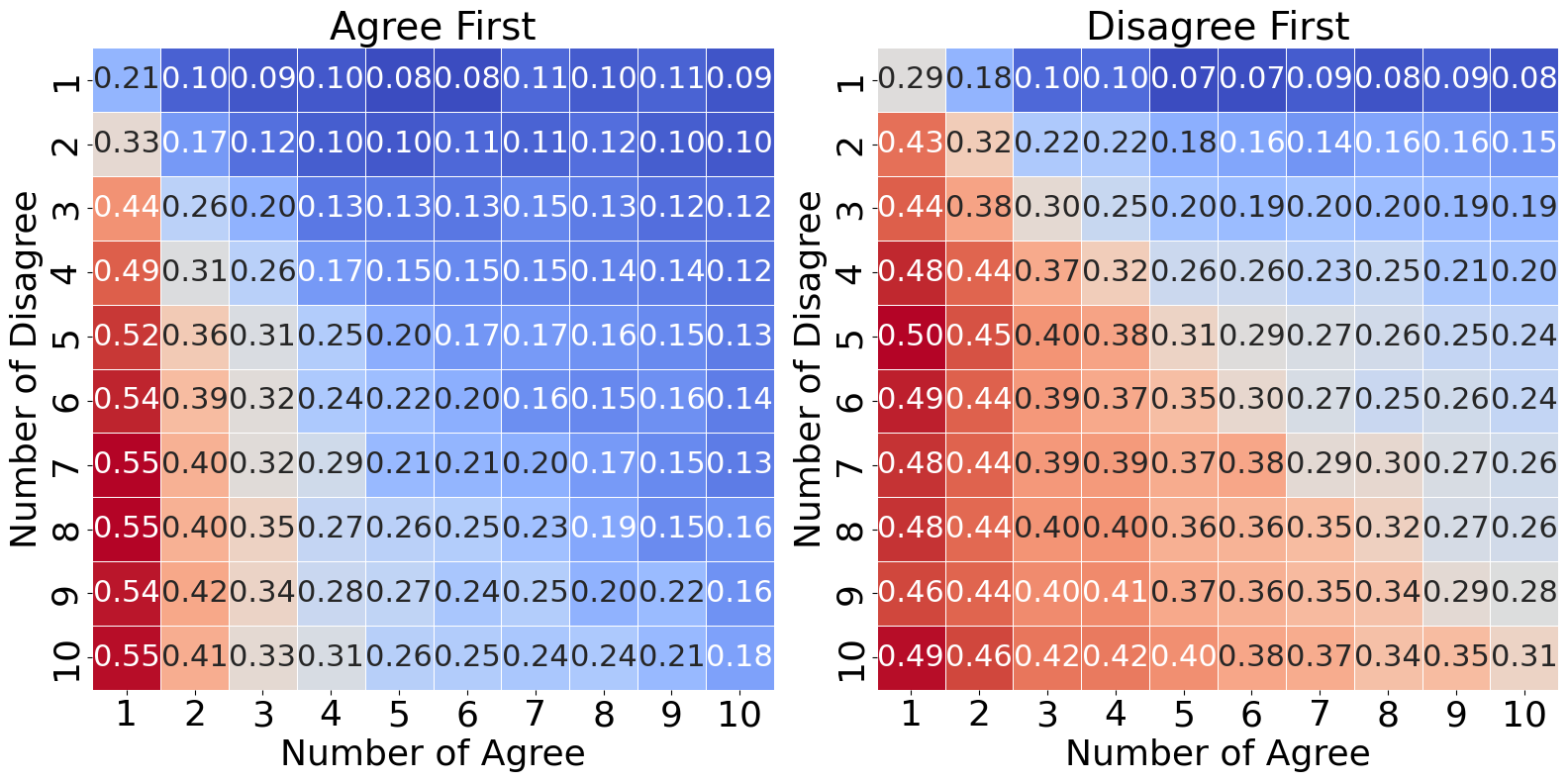}
    \caption{Comparison of average flip rates across two presentation orders. Each heatmap shows the average flip rate based on different combinations of agreeing and disagreeing peers. The x-axis represents the number of agreeing agents, and the y-axis represents the number of disagreeing agents. Higher flip rates are shown in red, while lower rates are shown in blue.}
    \label{fig:order}
\end{figure}

\section{Format of Peer Information - Modulator of Herd Behavior}

In complex collaborative settings involving large groups, confidence is shaped not only by the response content or peer demographics, but also by the number of peers who express agreement or disagreement with a given response. Prior research indicates that social validation, such as the quantity of agreeing peers, can strongly influence individual confidence, often exerting a greater effect than the intrinsic merit of the original response \cite{asch2016effects, moussaid2013social}.

The format in which peer information is conveyed to individuals is also crucial. In particular, how peer input is summarized and presented, especially when representing large groups, can shape perception in distinct ways. Furthermore, because peer information is communicated through language, its inherently sequential nature introduces an unavoidable ordering, which can influence how individuals interpret the information.

To explore the role of information format in affecting herd behavior, we conduct a series of experiments to assess how factors such as the number of agreeing or disagreeing agents, presentation methods, and the presentation order affect the magnitude of herd behavior.

\subsection{Experiment Setting}

We extended the experimental design from Section \ref{sec:confidence_setting} with several key modifications to examine the effects of peer information format on herd behavior.

\paragraph{I. Number of Agreeing and Disagreeing Agents}  
In contrast to the previous setup, which included only a single peer, we introduced multiple peers and categorized them into two groups: agreeing agents $A^A$ and disagreeing agents $A^D$. Agreeing agents share the same response as the target agent ($r_i = r_j$), while disagreeing agents provide the \textit{2nd} response type, the most persuasive alternative, as peer response.

\paragraph{II. Presentation Methods}  
To convey peer information to the target agent, we compared the following five methods of presentation:

    \noindent \textbf{• Count}: Present the number of peers supporting each response (e.g., "X agents think the answer is A").
    
    \noindent \textbf{• Ratio}: Present the percentage of peers supporting each response (e.g., "X\% of agents think the answer is A").
    
    \noindent \textbf{• List}: List the agents supporting each response (e.g., "Agents A, B, and C think the answer is A").
    
    \noindent \textbf{• Disc}: Display each peer's response individually (e.g., "Agent A thinks the answer is A; Agent B thinks the answer is B; ...").
    
    \noindent \textbf{• Reason}: Extend the \textit{Disc} method by including justifications for each response (e.g., "Agent A thinks the answer is A because ...").

\paragraph{III. Presentation Order}  
To assess the influence of order in information delivery, we employed two sequencing conditions: presenting agreeing agents ($A^A$) before disagreeing agents ($A^D$) (\textbf{Agree First}), and vice versa (\textbf{Disagree First}).

\subsection{Dataset}
We continue using the six benchmarks described in Section \ref{sec:confidence_dataset}, applying random sampling and capping the number of questions at 200 per benchmark to ensure data balance and adhere to budget constraints.

\subsection{Results}

\paragraph{(Dis)agreement Size and Herding} Table~\ref{tbl:format} and Figure~\ref{fig:format} illustrate how the strength of herding behavior varies with the size of agreement or disagreement. Specifically, Table~\ref{tbl:format} reports the average flip rate and the Pearson correlation between the flipping indicator $\mathbb{I}$ and the number of agreeing agents ($|A^A|$), disagreeing agents ($|A^D|$), and their difference ($|A^A| - |A^D|$), evaluated across different presentation formats and benchmark datasets. Overall, among the various formats, herding behavior is most pronounced when participants are presented with reasons. Interestingly, this effect is negligible on opinion-based benchmarks, suggesting that listing reasons is primarily effective in objective tasks. Moreover, across both factual and opinionated benchmarks, an increase in the number of agreeing agents or a decrease in the number of disagreeing agents generally leads to weaker herding behavior, and vice versa. Among all metrics, the difference between agreeing and disagreeing agents, reflecting the relative confidence between the individual and their peers, emerges as the strongest predictor of herding.

\paragraph{Effect of Presentation Format} Figure \ref{fig:format} compares five different presentation formats, where each subfigure displays a heatmap of the average flip rate as a function of the number of agreeing and disagreeing agents. The choice of presentation format significantly influences herding behavior. In the \textit{Count} and \textit{Ratio} formats, the heatmaps reveal a distinct separation into upper and lower triangles along the diagonal where the number of agreeing and disagreeing agents is equal. The upper triangle, representing cases where agreement outnumbers disagreement, exhibits consistently low flip rates regardless of the total number of peers, whereas the lower triangle shows much higher flip rates. This clear division suggests that numerical summaries of peer opinions help agents assess the balance between agreement and disagreement more effectively. In contrast, the other three formats, \textit{List}, \textit{Disc}, and \textit{Reason}, do not show such a sharp division. Instead, they demonstrate strong herding behavior only when the number of agreeing agents is small ($\leq 2$), and exhibit more resistance to change when three or more agents agree. Notably, the \textit{Reason} format results in the highest overall flip rates, even when agreement is greater, suggesting that providing justifications enhances persuasive power among agents.

\paragraph{Effect of Presentation Order} One notable finding from our experiment concerns the order in which information about agreeing and disagreeing agents is presented. Figure \ref{fig:order} displays heatmaps comparing two conditions: one where agreement is shown first, and another where disagreement is shown first, averaged across all presentation formats. The results reveal a strong difference between the two orders. When disagreement is presented first, herding behavior is generally stronger. The separation between the upper and lower triangles in the heatmap is more pronounced in this condition. In contrast, when agreement is shown first, high flip rates occur primarily when the number of agreeing agents is small ($\leq 2$), suggesting that the sequence in which peer opinions are revealed can influence tendency of herd behavior.

\begin{table*}[t]
\resizebox{\textwidth}{!}{%
\begin{tabular}{llllllll}
\hline
               & \multicolumn{4}{c}{MMLU-Pro}                                  & \multicolumn{3}{c}{GlobalOpinionQA}           \\
Condition      & Flip Rate (↑)     & Entropy (↓)      & Consensus Rate (↑)& Accuracy (↑)     & Flip Rate (↑)    & Entropy (↓)      & Consensus Rate (↑)\\ \hline
Original       & -             & 1.10          & 0.13            &  0.04               & -             & 0.82          &  0.14         \\
CoT       & -                  & 0.91          & 0.28            &  0.23               & -             & 0.63          &  0.34         \\
Baseline        & 0.55          & 0.43          & 0.56           &  0.18              & 0.44          & 0.29          &  0.69        \\
Strong Factors & \textbf{0.63*} & 0.43          & 0.54           &  \textbf{0.29*}     & \textbf{0.59*} & 0.49          & 0.46          \\
Weak Factors   & 0.36          & \textbf{0.28*} & \textbf{0.69*}  & 0.16               & 0.23          & \textbf{0.22*} & \textbf{0.76*}          \\
Strong Prompt  & 0.55          & 0.43          & 0.57            &  0.17               & 0.44          & 0.29          & 0.69          \\
Weak Prompt    & 0.55          & 0.43          & 0.56            &  0.18               & 0.46          & 0.36          & 0.61          \\ \hline
\end{tabular}
}
\caption{The effects of different control conditions on herd behaviors across factual and opinionated benchmarks. Flip rate, consensus rate, and accuracy (MMLU-Pro only) are higher-is-better metrics, while entropy is a lower-is-better metric. Bolded values are the best value in the column, and asterisks (*) denote statistical significance ($p < 0.05$) based on paired t-tests within each column. \textit{Strong Factors} yield the highest accuracy on MMLU-Pro and the highest flip rate on both datasets, indicating greater sensitivity to peer input. \textit{Weak Factors} exhibit the lowest entropy and highest consensus rate on both datasets, suggesting more aligned responses.}
\label{tbl:control}
\end{table*}

\section{Controllable Herd Behavior}

We have shown that herd behavior can be significantly influenced by factors such as presentation format, agreement size, and information order. These findings suggest that herd behavior is not fixed, but controllable.

In MAS applications, certain tasks benefit from strong herd behavior. For instance, in consensus-building or decision aggregation, quick convergence improves coordination and efficiency \cite{cho2024roundtable}. This is useful in tasks like collective prediction or distributed sensing. In contrast, tasks that rely on exploration or creativity, such as idea generation or strategy search, require diverse perspectives \cite{hong2023metagpt, xu2023language}. In these cases, strong herding can suppress innovation and lead to premature convergence, making independent reasoning more valuable.

Therefore, being able to control the strength of herd behavior is crucial. By adjusting how peer information is presented, we can encourage either convergence or independence depending on the task needs.

\subsection{Experiment Setting}

We simulate a collaborative scenario involving five agents. Each agent first generates an initial response to a given question $q$ using a high temperature ($\tau=1$) setting to promote diversity. Then, each agent independently revises their response after being shown the answers of the other four agents.The following metrics are used:

    \noindent \textbf{• Flip Rate}: Measures how often agents change their initial response.
    
    \noindent \textbf{• Entropy}: Quantifies the diversity in the final responses, reflecting overall alignment or disagreement among agents.
    
    \noindent \textbf{• Consensus Rate}: Indicates whether a unanimous consensus is reached.
    
    \noindent \textbf{• Accuracy}: For factual tasks, this captures the collective correctness of the agents' final responses. If a unanimous agreement is not reached, we mark it as incorrect. \newline

To assess the controllability of herd behavior, we compare the following conditions:

    \noindent \textbf{• Original}: Baseline condition using agents' initial responses before any peer input.

    \noindent \textbf{• CoT}: Extend the \textit{Original} condition by adding chain-of-thought reasoning \cite{wei2022chain}.
    
    \noindent \textbf{• Baseline}: Baseline condition without peer persona; both presentation format and order are randomized.
    
    \noindent \textbf{• Strong Factors}: Combines elements that amplify herding—\textit{graduate degree} persona, \textit{Reason} format, and showing disagreeing responses first.
    
    \noindent \textbf{• Weak Factors}: Combines elements that dampen herding—peers have a \textit{high school diploma}, use the \textit{Disc} format, and show agreeing responses first.
    
    \noindent \textbf{• Strong Prompt}: Uses the system prompt “Please be agreeable” to promote conformity, with random presentation format and order.
    
    \noindent \textbf{• Weak Prompt}: Uses the system prompt “Please be stubborn” to encourage resistance to peer influence, with random presentation format and order.

\subsection{Dataset}

Diversity in initial responses is essential for studying herd behavior. To ensure this, we filtered for questions where the highest probability among original responses was less than 0.8, indicating sufficient variation across agents. To maintain a reasonable dataset size, we selected two benchmarks: MMLU-Pro and GlobalOpinionQA, sampling 500 questions from each. It is worth noting that this filtering process favors more contentious or ambiguous questions, which may increase the difficulty of the task.

\subsection{Results}

\paragraph{Effect of Strong vs. Weak Factors on Herding} Table \ref{tbl:control} summarizes the impact of different control conditions on herd behavior across two benchmarks. The \textit{Strong Factors} condition yields the highest flip rate on both datasets (0.63 on MMLU-Pro and 0.59 on GlobalOpinionQA), indicating that agents are more likely to revise their answers when exposed to highly persuasive peer input. This setting also leads to the highest group accuracy (0.29) on MMLU-Pro, even higher than \textit{CoT}, suggesting that well-structured peer influence can improve collective performance on factual tasks. In contrast, the \textit{Weak Factors} condition results in the lowest flip rate (0.36 and 0.23) and entropy (0.28 and 0.22), demonstrating more consistent and aligned final responses with reduced peer influence. Despite reduced herding, the consensus rate remains high, suggesting that consensus can still emerge even when agents are less swayed.

\paragraph{Limited Effect of Prompt-Based Control} The \textit{Strong Prompt} and \textit{Weak Prompt} conditions show similar flip rates (0.55) and entropy levels (0.43) with \textit{Baseline}, indicating that prompt-level control has weaker effects compared to presentation factors, especially on the factual dataset (MMLU-Pro). While some effect is observed on the opinionated dataset, structural cues in peer presentation remain more effective in modulating herding.

\section{Discussion}
\subsection{Understanding and Controlling Herd Behavior}

Our findings reveal a nuanced picture of how herding emerges in multi-agent decision-making and the factors that modulate its intensity. Confidence alignment between the self and perceived peers plays a central role: herding is strongest when individuals feel uncertain while perceiving high confidence from peers. This dynamic is further shaped by social cues, where peer personas with higher status or domain relevance amplify conformity, particularly in objective tasks. However, this does not always lead to better outcomes; the drop in accuracy under the \textit{2nd} response type highlights the risks of misplaced trust. Furthermore, while herding is often viewed negatively, our results demonstrate that under carefully designed conditions, such as the \textit{Strong Factors} setting, peer input can enhance collective performance, suggesting that not all herding is detrimental.

Our study also underscores the importance of structural presentation in shaping social influence. Formats like \textit{Count} and \textit{Ratio} facilitate clear comparative reasoning, reducing flips when agreement is strong. Conversely, \textit{Reason} increases overall flip rates, emphasizing the persuasive power of justifications. Order of information presentation also matters: leading with disagreement encourages greater conformity than leading with agreement. Interestingly, prompt-level interventions had minimal effect compared to structural changes. Together, these insights offer actionable strategies for both harnessing and regulating herding behavior in collaborative AI and human-AI systems.

\subsection{(Ir)rationality in Agents}

Our analysis reveals that agents often behave rationally in response to confidence signals and social cues. Flip rates align with the interplay of self and perceived peer confidence: agents are more likely to switch when their own confidence is low and peers appear confident. Similarly, agents respond predictably to peer personas, with higher flip rates for authoritative figures. In \textit{Count} and \textit{Ratio} formats, flip behavior scales logically with the number of agreeing and disagreeing peers, suggesting quantitative reasoning based on social consensus.

However, we also observe deviations from rationality. Formats like \textit{List}, \textit{Disc}, and \textit{Reason} break the expected trend, showing weaker links between peer agreement size and flip rates. Presentation order also affects behavior, akin to first-impression bias, despite identical information. Moreover, prompt-based instructions have minimal effect compared to structural cues, indicating that agents are more influenced by framing than by explicit guidance. These findings point to bounded rationality shaped by presentation and context.

\section{Related Works}
Recent studies have explored the cognitive impacts and practical consequences of AI-driven systems, offering insights into how these technologies influence human reasoning and decision-making processes \cite{chen2024ai, shaki2023cognitive}. In parallel, research on the structural dynamics of language models has uncovered how architectural and training factors shape model behavior and outputs \cite{jumelet2024language, sinclair2022structural}. Additionally, a growing body of work has examined prosocial forms of irrationality, such as herd behavior, highlighting how collective decision-making can deviate from individual rationality while serving social cohesion or group benefits \cite{liu2024exploring}. However, these works have not thoroughly examined the underlying factors driving herd behavior or investigated the extent to which such behavior can be controlled.

\section{Conclusion}

This work presents a comprehensive analysis of herding behavior in multi-agent decision-making, revealing how confidence and presentation formats shape social influence. While agents often act rationally in response to structured cues, they remain vulnerable to framing effects and presentation biases. Our findings offer actionable insights for designing collaborative AI systems that balance influence and autonomy.  

\section*{Limitations}

While our study sheds light on the dynamics of herd behavior in LLM-based MAS, several limitations warrant discussion.

First, our experimental setup is constrained to controlled decision-making scenarios using multiple-choice questions across six benchmarks. Although these benchmarks span factual and opinionated domains, they may not fully capture the complexity and ambiguity of real-world collaborative tasks, such as open-ended discussions, multi-turn reasoning, or creative problem solving. The discrete nature of the response space may limit the generalizability of our findings to tasks requiring nuanced textual generation or longer context maintenance.

Second, we model perceived confidence and peer influence using static representations. These proxies may not capture the rich, dynamic interplay of trust, reputation, or credibility in more sophisticated agent interactions. Additionally, the absence of memory or learning mechanisms prevents agents from adapting their behavior over time, which could either dampen or exacerbate herd tendencies in longitudinal settings.

Third, our experiments involve agents from the same underlying language model architecture, which might limit behavioral diversity and obscure effects that could emerge from heterogeneous agents. Real-world MAS may involve agents with varying objectives, training data, or model sizes, introducing additional factors that could modulate conformity behaviors.

Finally, although we attempt to manipulate social influence through structured prompts and presentation formats, our findings on the weak efficacy of prompt-based controls suggest that LLMs may not reliably interpret meta-instructions in multi-agent settings. This points to a broader challenge in aligning emergent social behavior with high-level design intentions, particularly when using black-box models.

Future work could extend this research by incorporating more ecologically valid tasks, exploring heterogeneous agent configurations, and integrating adaptive learning mechanisms to better simulate evolving social dynamics in collaborative AI systems.

\bibliography{custom}

\begin{thebibliography}{37}
\providecommand{\natexlab}[1]{#1}

\bibitem[{Achiam et~al.(2023)Achiam, Adler, Agarwal, Ahmad, Akkaya, Aleman, Almeida, Altenschmidt, Altman, Anadkat et~al.}]{achiam2023gpt}
Josh Achiam, Steven Adler, Sandhini Agarwal, Lama Ahmad, Ilge Akkaya, Florencia~Leoni Aleman, Diogo Almeida, Janko Altenschmidt, Sam Altman, Shyamal Anadkat, et~al. 2023.
\newblock Gpt-4 technical report.
\newblock \emph{arXiv preprint arXiv:2303.08774}.

\bibitem[{Asch(2016)}]{asch2016effects}
Solomon~E Asch. 2016.
\newblock Effects of group pressure upon the modification and distortion of judgments.
\newblock In \emph{Organizational influence processes}, pages 295--303. Routledge.

\bibitem[{Banerjee(1992)}]{banerjee1992simple}
Abhijit~V Banerjee. 1992.
\newblock A simple model of herd behavior.
\newblock \emph{The quarterly journal of economics}, 107(3):797--817.

\bibitem[{Bang et~al.(2017)Bang, Aitchison, Moran, Herce~Castanon, Rafiee, Mahmoodi, Lau, Latham, Bahrami, and Summerfield}]{bang2017confidence}
Dan Bang, Laurence Aitchison, Rani Moran, Santiago Herce~Castanon, Banafsheh Rafiee, Ali Mahmoodi, Jennifer~YF Lau, Peter~E Latham, Bahador Bahrami, and Christopher Summerfield. 2017.
\newblock Confidence matching in group decision-making.
\newblock \emph{Nature Human Behaviour}, 1(6):0117.

\bibitem[{Bikhchandani et~al.(1992)Bikhchandani, Hirshleifer, and Welch}]{bikhchandani1992theory}
Sushil Bikhchandani, David Hirshleifer, and Ivo Welch. 1992.
\newblock A theory of fads, fashion, custom, and cultural change as informational cascades.
\newblock \emph{Journal of political Economy}, 100(5):992--1026.

\bibitem[{Chen et~al.(2024)Chen, Liu, Dong, Liu, Sakai, and Wu}]{chen2024ai}
Nuo Chen, Jiqun Liu, Xiaoyu Dong, Qijiong Liu, Tetsuya Sakai, and Xiao-Ming Wu. 2024.
\newblock Ai can be cognitively biased: An exploratory study on threshold priming in llm-based batch relevance assessment.
\newblock In \emph{Proceedings of the 2024 Annual International ACM SIGIR Conference on Research and Development in Information Retrieval in the Asia Pacific Region}, pages 54--63.

\bibitem[{Cho et~al.(2024)Cho, Shu, Das, Alkhouli, Lai, Cai, Sunkara, and Zhang}]{cho2024roundtable}
Young-Min Cho, Raphael Shu, Nilaksh Das, Tamer Alkhouli, Yi-An Lai, Jason Cai, Monica Sunkara, and Yi~Zhang. 2024.
\newblock Roundtable: Investigating group decision-making mechanism in multi-agent collaboration.
\newblock \emph{arXiv preprint arXiv:2411.07161}.

\bibitem[{Clark et~al.(2018)Clark, Cowhey, Etzioni, Khot, Sabharwal, Schoenick, and Tafjord}]{clark2018think}
Peter Clark, Isaac Cowhey, Oren Etzioni, Tushar Khot, Ashish Sabharwal, Carissa Schoenick, and Oyvind Tafjord. 2018.
\newblock Think you have solved question answering? try arc, the ai2 reasoning challenge.
\newblock \emph{arXiv preprint arXiv:1803.05457}.

\bibitem[{Du et~al.(2023)Du, Li, Torralba, Tenenbaum, and Mordatch}]{du2023improving}
Yilun Du, Shuang Li, Antonio Torralba, Joshua~B Tenenbaum, and Igor Mordatch. 2023.
\newblock Improving factuality and reasoning in language models through multiagent debate.
\newblock In \emph{Forty-first International Conference on Machine Learning}.

\bibitem[{Durmus et~al.(2023)Durmus, Nguyen, Liao, Schiefer, Askell, Bakhtin, Chen, Hatfield-Dodds, Hernandez, Joseph et~al.}]{durmus2023towards}
Esin Durmus, Karina Nguyen, Thomas~I Liao, Nicholas Schiefer, Amanda Askell, Anton Bakhtin, Carol Chen, Zac Hatfield-Dodds, Danny Hernandez, Nicholas Joseph, et~al. 2023.
\newblock Towards measuring the representation of subjective global opinions in language models.
\newblock \emph{arXiv preprint arXiv:2306.16388}.

\bibitem[{Fu et~al.(2017)Fu, Lee, and Danescu-Niculescu-Mizil}]{fu2017confidence}
Liye Fu, Lillian Lee, and Cristian Danescu-Niculescu-Mizil. 2017.
\newblock When confidence and competence collide: Effects on online decision-making discussions.
\newblock In \emph{Proceedings of the 26th international conference on world wide web}, pages 1381--1390.

\bibitem[{Guo et~al.(2024)Guo, Chen, Wang, Chang, Pei, Chawla, Wiest, and Zhang}]{guo2024large}
Taicheng Guo, Xiuying Chen, Yaqi Wang, Ruidi Chang, Shichao Pei, Nitesh~V Chawla, Olaf Wiest, and Xiangliang Zhang. 2024.
\newblock Large language model based multi-agents: A survey of progress and challenges.
\newblock \emph{arXiv preprint arXiv:2402.01680}.

\bibitem[{Hong et~al.(2023)Hong, Zheng, Chen, Cheng, Wang, Zhang, Wang, Yau, Lin, Zhou et~al.}]{hong2023metagpt}
Sirui Hong, Xiawu Zheng, Jonathan Chen, Yuheng Cheng, Jinlin Wang, Ceyao Zhang, Zili Wang, Steven Ka~Shing Yau, Zijuan Lin, Liyang Zhou, et~al. 2023.
\newblock Metagpt: Meta programming for multi-agent collaborative framework.
\newblock \emph{arXiv preprint arXiv:2308.00352}, 3(4):6.

\bibitem[{Jumelet et~al.(2024)Jumelet, Zuidema, and Sinclair}]{jumelet2024language}
Jaap Jumelet, Willem Zuidema, and Arabella Sinclair. 2024.
\newblock Do language models exhibit human-like structural priming effects?
\newblock \emph{arXiv preprint arXiv:2406.04847}.

\bibitem[{Laban et~al.(2023)Laban, Murakhovs'~ka, Xiong, and Wu}]{laban2023you}
Philippe Laban, Lidiya Murakhovs'~ka, Caiming Xiong, and Chien-Sheng Wu. 2023.
\newblock Are you sure? challenging llms leads to performance drops in the flipflop experiment.
\newblock \emph{arXiv preprint arXiv:2311.08596}.

\bibitem[{Liu et~al.(2024{\natexlab{a}})Liu, Wang, Huang, Xu, Zeng, Jiang, Yang, and Li}]{liu2024groupdebate}
Tongxuan Liu, Xingyu Wang, Weizhe Huang, Wenjiang Xu, Yuting Zeng, Lei Jiang, Hailong Yang, and Jing Li. 2024{\natexlab{a}}.
\newblock Groupdebate: Enhancing the efficiency of multi-agent debate using group discussion.
\newblock \emph{arXiv preprint arXiv:2409.14051}.

\bibitem[{Liu et~al.(2024{\natexlab{b}})Liu, Zhang, Shang, Guo, Yang, and Zhu}]{liu2024exploring}
Xuan Liu, Jie Zhang, Haoyang Shang, Song Guo, Chengxu Yang, and Quanyan Zhu. 2024{\natexlab{b}}.
\newblock Exploring prosocial irrationality for llm agents: A social cognition view.
\newblock \emph{arXiv preprint arXiv:2405.14744}.

\bibitem[{Moussa{\"\i}d et~al.(2013)Moussa{\"\i}d, K{\"a}mmer, Analytis, and Neth}]{moussaid2013social}
Mehdi Moussa{\"\i}d, Juliane~E K{\"a}mmer, Pantelis~P Analytis, and Hansj{\"o}rg Neth. 2013.
\newblock Social influence and the collective dynamics of opinion formation.
\newblock \emph{PloS one}, 8(11):e78433.

\bibitem[{Muchnik et~al.(2013)Muchnik, Aral, and Taylor}]{muchnik2013social}
Lev Muchnik, Sinan Aral, and Sean~J Taylor. 2013.
\newblock Social influence bias: A randomized experiment.
\newblock \emph{Science}, 341(6146):647--651.

\bibitem[{OpenAI(2023)}]{openai2023gpt4omini}
OpenAI. 2023.
\newblock \href {https://openai.com/index/gpt-4o-mini-advancing-cost-efficient-intelligence/} {Gpt-4o-mini: Advancing cost-efficient intelligence}.
\newblock Accessed: 2024-08-18.

\bibitem[{{OpenAI}(2024)}]{openai2024gpt41}
{OpenAI}. 2024.
\newblock {GPT-4.1}.
\newblock \url{https://openai.com/index/gpt-4-1/}.
\newblock Accessed: 2025-05-20.

\bibitem[{Park et~al.(2023)Park, O'Brien, Cai, Morris, Liang, and Bernstein}]{park2023generative}
Joon~Sung Park, Joseph O'Brien, Carrie~Jun Cai, Meredith~Ringel Morris, Percy Liang, and Michael~S Bernstein. 2023.
\newblock Generative agents: Interactive simulacra of human behavior.
\newblock In \emph{Proceedings of the 36th annual acm symposium on user interface software and technology}, pages 1--22.

\bibitem[{Pescetelli et~al.(2021)Pescetelli, Hauperich, and Yeung}]{pescetelli2021confidence}
Niccol{\`o} Pescetelli, Anna-Katharina Hauperich, and Nick Yeung. 2021.
\newblock Confidence, advice seeking and changes of mind in decision making.
\newblock \emph{Cognition}, 215:104810.

\bibitem[{Raafat et~al.(2009)Raafat, Chater, and Frith}]{raafat2009herding}
Ramsey~M Raafat, Nick Chater, and Chris Frith. 2009.
\newblock Herding in humans.
\newblock \emph{Trends in cognitive sciences}, 13(10):420--428.

\bibitem[{Rein et~al.(2024)Rein, Hou, Stickland, Petty, Pang, Dirani, Michael, and Bowman}]{rein2024gpqa}
David Rein, Betty~Li Hou, Asa~Cooper Stickland, Jackson Petty, Richard~Yuanzhe Pang, Julien Dirani, Julian Michael, and Samuel~R Bowman. 2024.
\newblock Gpqa: A graduate-level google-proof q\&a benchmark.
\newblock In \emph{First Conference on Language Modeling}.

\bibitem[{Santurkar et~al.(2023)Santurkar, Durmus, Ladhak, Lee, Liang, and Hashimoto}]{santurkar2023whose}
Shibani Santurkar, Esin Durmus, Faisal Ladhak, Cinoo Lee, Percy Liang, and Tatsunori Hashimoto. 2023.
\newblock Whose opinions do language models reflect?
\newblock In \emph{International Conference on Machine Learning}, pages 29971--30004. PMLR.

\bibitem[{Sap et~al.(2019)Sap, Rashkin, Chen, LeBras, and Choi}]{sap2019socialiqa}
Maarten Sap, Hannah Rashkin, Derek Chen, Ronan LeBras, and Yejin Choi. 2019.
\newblock Socialiqa: Commonsense reasoning about social interactions.
\newblock \emph{arXiv preprint arXiv:1904.09728}.

\bibitem[{Shaki et~al.(2023)Shaki, Kraus, and Wooldridge}]{shaki2023cognitive}
Jonathan Shaki, Sarit Kraus, and Michael Wooldridge. 2023.
\newblock Cognitive effects in large language models.
\newblock In \emph{ECAI 2023}, pages 2105--2112. IOS Press.

\bibitem[{Sinclair et~al.(2022)Sinclair, Jumelet, Zuidema, and Fern{\'a}ndez}]{sinclair2022structural}
Arabella Sinclair, Jaap Jumelet, Willem Zuidema, and Raquel Fern{\'a}ndez. 2022.
\newblock Structural persistence in language models: Priming as a window into abstract language representations.
\newblock \emph{Transactions of the Association for Computational Linguistics}, 10:1031--1050.

\bibitem[{Wang et~al.(2024)Wang, Ma, Zhang, Ni, Chandra, Guo, Ren, Arulraj, He, Jiang et~al.}]{wang2024mmlu}
Yubo Wang, Xueguang Ma, Ge~Zhang, Yuansheng Ni, Abhranil Chandra, Shiguang Guo, Weiming Ren, Aaran Arulraj, Xuan He, Ziyan Jiang, et~al. 2024.
\newblock Mmlu-pro: A more robust and challenging multi-task language understanding benchmark.
\newblock In \emph{The Thirty-eight Conference on Neural Information Processing Systems Datasets and Benchmarks Track}.

\bibitem[{Wei et~al.(2022)Wei, Wang, Schuurmans, Bosma, Xia, Chi, Le, Zhou et~al.}]{wei2022chain}
Jason Wei, Xuezhi Wang, Dale Schuurmans, Maarten Bosma, Fei Xia, Ed~Chi, Quoc~V Le, Denny Zhou, et~al. 2022.
\newblock Chain-of-thought prompting elicits reasoning in large language models.
\newblock \emph{Advances in neural information processing systems}, 35:24824--24837.

\bibitem[{Weng et~al.(2025)Weng, Chen, and Wang}]{weng2025we}
Zhiyuan Weng, Guikun Chen, and Wenguan Wang. 2025.
\newblock Do as we do, not as you think: the conformity of large language models.
\newblock \emph{arXiv preprint arXiv:2501.13381}.

\bibitem[{Wu and Ito(2025)}]{wu2025hidden}
Zengqing Wu and Takayuki Ito. 2025.
\newblock The hidden strength of disagreement: Unraveling the consensus-diversity tradeoff in adaptive multi-agent systems.
\newblock \emph{arXiv preprint arXiv:2502.16565}.

\bibitem[{Xiao and Wang(2019)}]{xiao2019quantifying}
Yijun Xiao and William~Yang Wang. 2019.
\newblock Quantifying uncertainties in natural language processing tasks.
\newblock In \emph{Proceedings of the AAAI conference on artificial intelligence}, volume~33, pages 7322--7329.

\bibitem[{Xu et~al.(2023)Xu, Yu, Fang, Wang, and Wu}]{xu2023language}
Zelai Xu, Chao Yu, Fei Fang, Yu~Wang, and Yi~Wu. 2023.
\newblock Language agents with reinforcement learning for strategic play in the werewolf game.
\newblock \emph{arXiv preprint arXiv:2310.18940}.

\bibitem[{Zarnoth and Sniezek(1997)}]{zarnoth1997social}
Paul Zarnoth and Janet~A Sniezek. 1997.
\newblock The social influence of confidence in group decision making.
\newblock \emph{Journal of Experimental Social Psychology}, 33(4):345--366.

\bibitem[{Zhu et~al.(2024)Zhu, Zhang, Stafford, Collier, and Vlachos}]{zhu2024conformity}
Xiaochen Zhu, Caiqi Zhang, Tom Stafford, Nigel Collier, and Andreas Vlachos. 2024.
\newblock Conformity in large language models.
\newblock \emph{arXiv preprint arXiv:2410.12428}.

\end{thebibliography}

\appendix
\section{Different LLMs}
\begin{table*}[t]
\resizebox{\textwidth}{!}{%
\begin{tabular}{lllllllll}
\hline
\textbf{}                                                            & \multicolumn{4}{c}{\textbf{Factual}}                                                                                 & \multicolumn{4}{c}{\textbf{Opinionated}}                                                                          \\
\textbf{\begin{tabular}[c]{@{}l@{}}Presentation\\ Format\end{tabular}} & \textbf{\begin{tabular}[c]{@{}l@{}}Avg. \\ Flip Rate\end{tabular}} & \textbf{$\rho(\mathbb{I},|A^A|)$} & \textbf{$\rho(\mathbb{I},|A^D|)$} & \textbf{$\rho(\mathbb{I},|A^A|-|A^D|)$} & \textbf{\begin{tabular}[c]{@{}l@{}}Avg. \\ Flip Rate\end{tabular}} & \textbf{$\rho(\mathbb{I},|A^A|)$} & \textbf{$\rho(\mathbb{I},|A^D|)$} & \textbf{$\rho(\mathbb{I},|A^A|-|A^D|)$} \\ \hline
Count                                                                & 0.22                                                               & -0.17           & 0.16         & -0.23          & 0.27                                                               & -0.36        & 0.31         & -0.47          \\
Ratio                                                                & 0.22                                                               & -0.28           & 0.25         & -0.37          & 0.30                                                               & -0.41        & 0.38         & -0.56          \\
List                                                                 & 0.21                                                               & -0.17           & 0.15         & -0.23          & 0.30                                                               & -0.22        & 0.22         & -0.31          \\
Disc                                                                 & 0.21                                                               & -0.11           & 0.09         & -0.14          & 0.28                                                               & -0.24        & 0.23         & -0.33          \\
Reason                                                               & 0.30                                                               & -0.05           & 0.05         & -0.07          & 0.30                                                               & -0.11        & 0.11         & -0.15          \\ \hline
\end{tabular}
}
\caption{Average flip rate and Pearson r correlation between the flipping indicator $\mathbb{I}$ and the number of agreeing agents ($|A^A|$), disagreeing agents ($|A^D|$), or their difference ($|A^A| - |A^D|$), evaluated across various presentation formats and benchmark datasets. All Pearson r correlations are statistically significant (p < 0.001). Overall, herd behavior is strongest when presented with reasons. The difference between agreeing and disagreeing agents is the strongest predictor of herd behavior.}
\label{tbl:format}
\end{table*}
\begin{table*}[t]
\resizebox{\textwidth}{!}{%
\begin{tabular}{llllllll}
\hline
\multirow{2}{*}{\textbf{Method}} & \multicolumn{3}{c}{\textbf{Factual}}                                                                                                                                                                & \multicolumn{3}{c}{\textbf{Opinionated}}                            & \multirow{2}{*}{\textbf{Average}} \\
                                 & \textbf{\begin{tabular}[c]{@{}l@{}}MMLU-\\ Pro\end{tabular}} & \textbf{\begin{tabular}[c]{@{}l@{}}GPQA-\\ Diamond\end{tabular}} & \textbf{\begin{tabular}[c]{@{}l@{}}ARC-\\ Challenge\end{tabular}} & \textbf{OpinionQA} & \textbf{GlobalOpinionQA} & \textbf{SOCIAL IQA} &                                   \\ \hline
gpt-4o-mini\_2nd                 & \textbf{0.5*}                                                & \textbf{0.59*}                                                   & \textbf{0.11*}                                                    & \textbf{0.64*}     & \textbf{0.71*}           & \textbf{0.16*}      & \textbf{0.45*}                    \\
gpt-4o-mini\_last                & 0.25                                                         & 0.42                                                             & 0.06                                                              & 0.54               & 0.59                     & 0.06                & 0.32                              \\
gpt-4o\_2nd                      & \textbf{0.55*}                                               & \textbf{0.70*}                                                   & \textbf{0.06*}                                                    & \textbf{0.52*}     & \textbf{0.57*}           & \textbf{0.26*}      & \textbf{0.44*}                    \\
gpt-4o\_last                     & 0.34                                                         & 0.45                                                             & 0.02                                                              & 0.34               & 0.43                     & 0.14                & 0.29                              \\
gpt-4.1\_2nd                     & \textbf{0.51*}                                               & \textbf{0.63*}                                                   & \textbf{0.05*}                                                    & \textbf{0.66*}     & \textbf{0.66*}           & \textbf{0.22*}      & \textbf{0.45*}                    \\
gpt-4.1\_last                    & 0.23                                                         & 0.35                                                             & 0.02                                                              & 0.49               & 0.54                     & 0.10                & 0.29                              \\
gpt-4.1-mini\_2nd                & \textbf{0.46*}                                               & \textbf{0.49*}                                                   & \textbf{0.04*}                                                    & \textbf{0.4*}      & \textbf{0.42*}           & \textbf{0.2*}       & \textbf{0.33*}                    \\
gpt-4.1-mini\_last               & 0.30                                                         & 0.32                                                             & 0.00                                                              & 0.25               & 0.33                     & 0.11                & 0.22                              \\
gpt-4.1-nano\_2nd                & \textbf{0.62*}                                               & \textbf{0.57*}                                                   & \textbf{0.18*}                                                    & \textbf{0.51*}     & \textbf{0.62*}           & \textbf{0.33*}      & \textbf{0.47*}                    \\
gpt-4.1-nano\_last               & 0.46                                                         & 0.39                                                             & 0.12                                                              & 0.42               & 0.52                     & 0.16                & 0.34                              \\ \hline
\end{tabular}
}
\caption{Flip rates for \textit{2nd} and \textit{last} response types across different LLMs, used to assess the generalizability of perceived confidence effects on herd behavior. Bolded values indicate the highest flip rate within each group, reflecting the greatest herd influence. Asterisks (*) mark statistically significant differences (p < 0.05) based on paired t-tests conducted within each group.}
\label{tbl:llms}
\end{table*}
In our experiments, we employ \texttt{gpt-4o-mini-2024-07-18} \cite{openai2023gpt4omini}, using all default settings except for the temperature parameter, which is set to 0 unless stated otherwise.

In this section, we report flip rates across different peer conditions and datasets using a range of LLMs, including \texttt{gpt-4o (2024-11-20)}, \texttt{gpt-4o-mini (2024-07-18)}, \texttt{gpt-4.1 (2025-04-14)}, \texttt{gpt-4.1-mini (2025-04-14)}, and \texttt{gpt-4.1-nano (2025-04-14)} \cite{achiam2023gpt, openai2024gpt41}. We compare models' responses when the perceived confidence occurs at the \textit{2nd} versus the \textit{last} position. Table~\ref{tbl:llms} summarizes flip rates across all models and datasets. Overall, responses positioned \textit{2nd} consistently exhibit higher flip rates than those in the \textit{last} position, suggesting a greater susceptibility to herd behavior when confidence is perceived earlier.

\begin{table}[t]
\resizebox{\linewidth}{!}{%
\begin{tabular}{lllll}
\hline
                                          & \multicolumn{3}{c}{\textbf{Factual}}                                                                                                                                                                &                                    \\
\multirow{-2}{*}{\textbf{Peer Condition}} & \textbf{\begin{tabular}[c]{@{}l@{}}MMLU-\\ Pro\end{tabular}} & \textbf{\begin{tabular}[c]{@{}l@{}}GPQA-\\ Diamond\end{tabular}} & \textbf{\begin{tabular}[c]{@{}l@{}}ARC-\\ Challenge\end{tabular}} & \multirow{-2}{*}{\textbf{Average}} \\ \hline
\rowcolor[HTML]{C0C0C0} 
Original                                  & 0.45                                                         & 0.35                                                             & 0.93                                                              & 0.49                               \\
1st                                       & 0.45                                                         & 0.34                                                             & 0.92                                                              & 0.49                               \\
2nd                                       & \textbf{0.41*}                                               & \textbf{0.30}                                                    & \textbf{0.90*}                                                    & \textbf{0.45*}                     \\
rnd                                       & 0.42                                                         & 0.32                                                             & 0.91                                                              & 0.46                               \\
last                                      & 0.43                                                         & 0.35                                                             & 0.91                                                              & 0.48                               \\ \hline
Graduate Degree                           & \textbf{0.40*}                                               & \textbf{0.29}                                                    & \textbf{0.89}                                                     & \textbf{0.44*}                     \\
College Degree                            & 0.41                                                         & 0.32                                                             & 0.90                                                              & 0.45                               \\
High School Diploma                       & 0.41                                                         & 0.31                                                             & 0.90                                                              & 0.45                               \\ \hline
Employer                                  & \textbf{0.40*}                                               & 0.27                                                             & \textbf{0.74*}                                                    & \textbf{0.44*}                     \\
Employee                                  & 0.41                                                         & 0.27                                                             & 0.76                                                              & 0.45                               \\ \hline
In-Domain                                 & \textbf{0.40*}                                               & 0.29                                                             & -                                                                 & 0.40                               \\
Out-Of-Domain                             & 0.41                                                         & 0.29                                                             & -                                                                 & 0.40                               \\ \hline
\end{tabular}
}
\caption{Accuracy of revised response after receiving peer information. The first row, Original, represents accuracy of original response before receiving peer information. Bolded values represent the lowest accuracy within each group, and asterisks (*) denote statistical significance (p < 0.05) based on paired t-tests within each group.}
\label{tbl:accuracy}
\end{table}

\section{Prompt Details}
In this section, we present the prompt we use in our experiment settings. 

Below are persona prompt used for peer education:

\begin{tcolorbox}[title=\footnotesize Peer Education,top=1mm,bottom=1mm]
\scriptsize
Graduate Degree: \\
\texttt{Agent A's highest degree is graduate degree. Agent A thinks the answer is \{$r_j$\}.}\\

College Degree: \\
\texttt{Agent A's highest degree is college degree. Agent A thinks the answer is \{$r_j$\}.}\\

High School Diploma:\\
\texttt{Agent A's highest degree is high school diploma. Agent A thinks the answer is \{$r_j$\}.}\\
\end{tcolorbox}

Below are persona prompt used for peer social hierarchy:

\begin{tcolorbox}[title=\footnotesize Peer Social Hierarchy,top=1mm,bottom=1mm]
\scriptsize
Employer: \\
\texttt{Your employer thinks the answer is \{$r_j$\}.}\\

Employee: \\
\texttt{Your employee thinks the answer is \{$r_j$\}.}\\
\end{tcolorbox}

Below is persona prompt used for domain expertise. There are 14 domains in MMLU-Pro (\textit{law, computer science, history, health, philosophy, psychology, business, engineering, biology, chemistry, physics, math, economics, other}), and 3 domains in GPQA-Diamond (\textit{Biology, Physics, Chemistry}). While in-domain examples gives the same domain as the given question, out-domain ramdomly picked one from the complement set.

\begin{tcolorbox}[title=\footnotesize Peer Domain Expertice,top=1mm,bottom=1mm]
\scriptsize
\texttt{Agent A is an expert in \{$domain$\} domain. Agent A thinks the answer is \{$r_j$\}.}\\
\end{tcolorbox}

Below are prompt used for presentation methods:

\begin{tcolorbox}[title=\footnotesize Presentation Methods,top=1mm,bottom=1mm]
\scriptsize
Count:\\
\texttt{\{$agree\_size$\} agent\{$plural$\} think\{$s$\} the answer is \{$r_j$\}.}\\
and vise versa to disagreeing agents.\\

Ratio:\\
\texttt{Among \{$peer\_size$\} agents,\\
\{$agree\_ratio$\}\% think the answer is \{$r_j$\}.}\\
and vise versa to disagreeing agents.\\

List:\\
\texttt{Agent \{$list\_of\_agree\_agents$\} think the answer is \{$r_j$\}.\\
Agent \{$list\_of\_disagree\_agents$\} think the answer is \{$r_k$\}.}\\

Disc:\\
\texttt{Agent A think the answer is \{$r_j$\}.\\
Agent B think the answer is \{$r_k$\}.\\
...}\\

Reason:\\
\texttt{Agent A think the answer is \{$r_j$\}, because \{$reason_j$\}.\\
Agent B think the answer is \{$r_k$\}, because \{$reason_k$\}.\\
...}\\

\end{tcolorbox}

\section{Details of Datasets}
MMLU-Pro, GPQA-Diamond, ARC-Challenge is under MIT license, ARC-Challenge is under cc-by-sa-4.0 license, OpinionQA and SOCIAL IQA is not under a license, and GlobalOpinionQA is under cc-by-nc-sa-4.0 license. Our use of the dataset is consistent with the intended use. The datasets do not contain personally identifying info or offensive content. All the datasets are in english.

\end{document}